\documentclass[aps,twocolumn,amsmath,amssymb,prd,preprintnumbers,superscriptaddress,nofootinbib]{revtex4}

\usepackage{amsmath}
\usepackage{bm}
\usepackage[dvips]{graphicx}
\usepackage{color}
\newcommand{\be}{\begin{equation}}

\newcommand{\ee}{\end{equation}}
\newcommand{\bea}{\begin{eqnarray}}
\newcommand{\eea}{\end{eqnarray}}

\begin{document}
\preprint{BARI-TH 576/07}

\author{R.~Anglani}\email{roberto.anglani@ba.infn.it}
\affiliation{Dipartimento di Fisica,
Universit\`a di Bari, I-70126 Bari, Italia}\affiliation{I.N.F.N.,
Sezione di Bari, I-70126 Bari, Italia}\affiliation{Institute of Theoretical Physics, University of Wroclaw, pl.~Maksa Borna 9,
PL-50204 Wroclaw, Poland}

\author{M.~Ciminale}\email{marco.ciminale@ba.infn.it}
\affiliation{Dipartimento di Fisica, Universit\`a di Bari, I-70126
Bari, Italia}\affiliation{I.N.F.N., Sezione di Bari, I-70126 Bari,
Italia}

\author{N.~D.~Ippolito}\email{nicola.ippolito@ba.infn.it}
\affiliation{Dipartimento di Fisica, Universit\`a di Bari, I-70126
Bari, Italia}\affiliation{I.N.F.N., Sezione di Bari, I-70126 Bari,
Italia}

\date{\today}

\title{Recent advances in the three flavor Larkin-Ovchinnikov-Fulde-Ferrell phase of QCD}

\begin{abstract}
 We present a summary of the recent advances achieved in the study of three flavor Larkin-Ovchinnikov-Fulde-Ferrell (LOFF)
  phase of QCD. We have explored, using a Ginzburg-Landau expansion of the free energy, the LOFF phase with three flavors,
 in the simplest single plane wave structure, using the NJL four-fermion coupling. We have found that this phase does not
suffer the chromo-magnetic instability problem. A preliminary study
of astrophysical effects of quark matter in the aforementioned phase
has been done and we have evaluated self-consistently the strange
quark mass extending the pairing ansatz to the CubeX and 2Cube45z
structure. Finally we have investigated the possibility of Goldstone
bosons condensation
 in the favored cubic structures of LOFF phase.
\end{abstract}

\maketitle

%%%%%%%%%%%%%%%%%%%%%%%%%%%%%%%%%%%%%%%%%%%%
%% MAINMATTER
%%%%%%%%%%%%%%%%%%%%%%%%%%%%%%%%%%%%%%%%%%%%

\section{Introduction}

 The comprehension of the structure of the QCD phase diagram is one
of the most challenging topics within the elementary particle
physics. In particular, the study of the region corresponding to
very low temperatures and high densities (roughly from 3 to 10 times
the nuclear saturation density) had a big impulse some years ago
when a Cooper pairing among quarks, driven by the strong
interaction, was hypothesized to yield a collective phenomenon named
Color Superconductivity (CSC) \cite{alford}. At asymptotical
densities, the ground state of quark matter is successfully
described by the energetically favored phase named
Color-Flavor-Locking (CFL) \cite{CFL} in which all the light quarks
$u$, $d$, $s$ of any color form Cooper pairs with zero total
momentum and all fermionic excitations are gapped. At intermediate
densities, where it is not possible to neglect the strange quark
mass and the mismatch $\delta\mu$ in the quark chemical potentials
due to $\beta$-equilibrium and color and electrical neutrality, the
situation is much more involved. The ground state of matter in these
conditions is still matter of debate and several possible states
have been suggested. In particular the superconductive phases
characterized by gapless fermionic excitations, gapless-2SC (g2SC)
\cite{g2SC} and gapless-CFL (gCFL) \cite{gCFL} have been widely
discussed. However, it has been shown that these gapless phases
suffer chromomagnetic instability \cite{chromo} due to the imaginary
Meissner masses of some of the gluons associated with broken gauge
symmetries (an instability is present also in 2SC phase
\cite{chromo-2SC}).

Another possible phase largely discussed in literature is the Larkin-Ovchinnikov-Fulde-Ferrell (LOFF) phase \cite{LOFF-origin}. This phase is relevant since for appropriate values of $\delta\mu$, it can be energetically favored to form Cooper pairs with non-vanishing total momentum $\mathbf{p_{1}}+\mathbf{p_{2}}=2\mathbf{q}\neq0$, see \cite{LOFF-2} and for a review \cite{LOFF-review}. The two flavor LOFF phase has been found energetically favored with respect to the 2SC phase \cite{Giannakis1} and characterized by real gluon Meissner masses \cite{Giannakis2}. Anyway a more realistic description of QCD at intermediates densities requires that all the three quarks $u$, $d$ and $s$ should be taken into account, leading to a theoretical study of the three flavor LOFF phase of QCD.

In connection to this necessity we present, in the following, some important phenomenological results of the recent theoretical studies concerning the three flavor LOFF phase.

\section{Three flavor LOFF phase with single plane wave structure}
The first work, whose results we would like to show, about the three
flavor LOFF phase \cite{Casalbuoni:2005zp}, considered the following
ansatz for the spatial dependence of the order parameter:
\begin{equation}
<\psi_{i\alpha}\,C\,\gamma_5\,\psi_{\beta j}> =
\sum_{I=1}^{3}\,\Delta_I({\bf r})\,\epsilon^{\alpha\beta
I}\,\epsilon_{ijI}~\label{cond}
\end{equation}with
\begin{equation} \Delta_I ({\bf r}) = \Delta_I \exp\left(2i\,{\bf q_I}\cdot{\bf
r}\right)~. \label{eq:1Ws}
\end{equation}
The three independent functions $\Delta_{1}({\bf r})$,
$\Delta_{2}({\bf r})$, $\Delta_{3}({\bf r})$ describe respectively
$d-s$, $u-s$ and $u-d$ pairing. The same convention holds for the
wave vectors ${\bf q_I}$.  This means that all the pairs choose the
same direction for their wave vectors, so being all parallel or
antiparallel relative to one another.

The results of \cite{Casalbuoni:2005zp} show that there is a window
$128\; {\rm MeV} < m^2_s/\mu < 150\; {\rm MeV}$ where the LOFF state
has a lower free energy with respect to the normal phase and
homogeneous color-superconductive phases. The energetically favored
structure has $\Delta_2=\Delta_3$ and $\Delta_1=0$, and $\bf q_2$,
$\bf q_3$ parallel. The study is performed within a Ginzburg-Landau
expansion of the free energy up to the $\Delta^4$ order. The
interaction between quarks is considered as a NJL four-fermion
coupling in the mean field approximation. The strange quark mass is
treated as a shift of the corresponding chemical potential:
$\mu_s\rightarrow\mu_s-m_s^2/(2\mu)$. Finally the
$\beta$-equilibrium and color and electrical neutrality have been
imposed. The validity of the GL analysis with the above
approximations has been confirmed in \cite{Mannarelli:2006fy}.

\subsection{Chromomagnetic stability of the LOFF phase} The introduction of
a non-zero strange quark mass, together with the aforementioned
neutrality conditions, leads to a thermodynamic system of quarks
with different chemical potentials, and this in turns gives
gapless dispersion laws. This was first observed in other
(homogeneous) color superconducting phases, like gCFL and g2SC,
and in both cases it was proved to drive the so-called
chromo-magnetic instability of the system, meaning that the
screening Meissner masses of some gluons are imaginary
\cite{Casalbuoni:2004tb}.

\begin{figure}[h!] \centering
{\includegraphics[width=7cm]{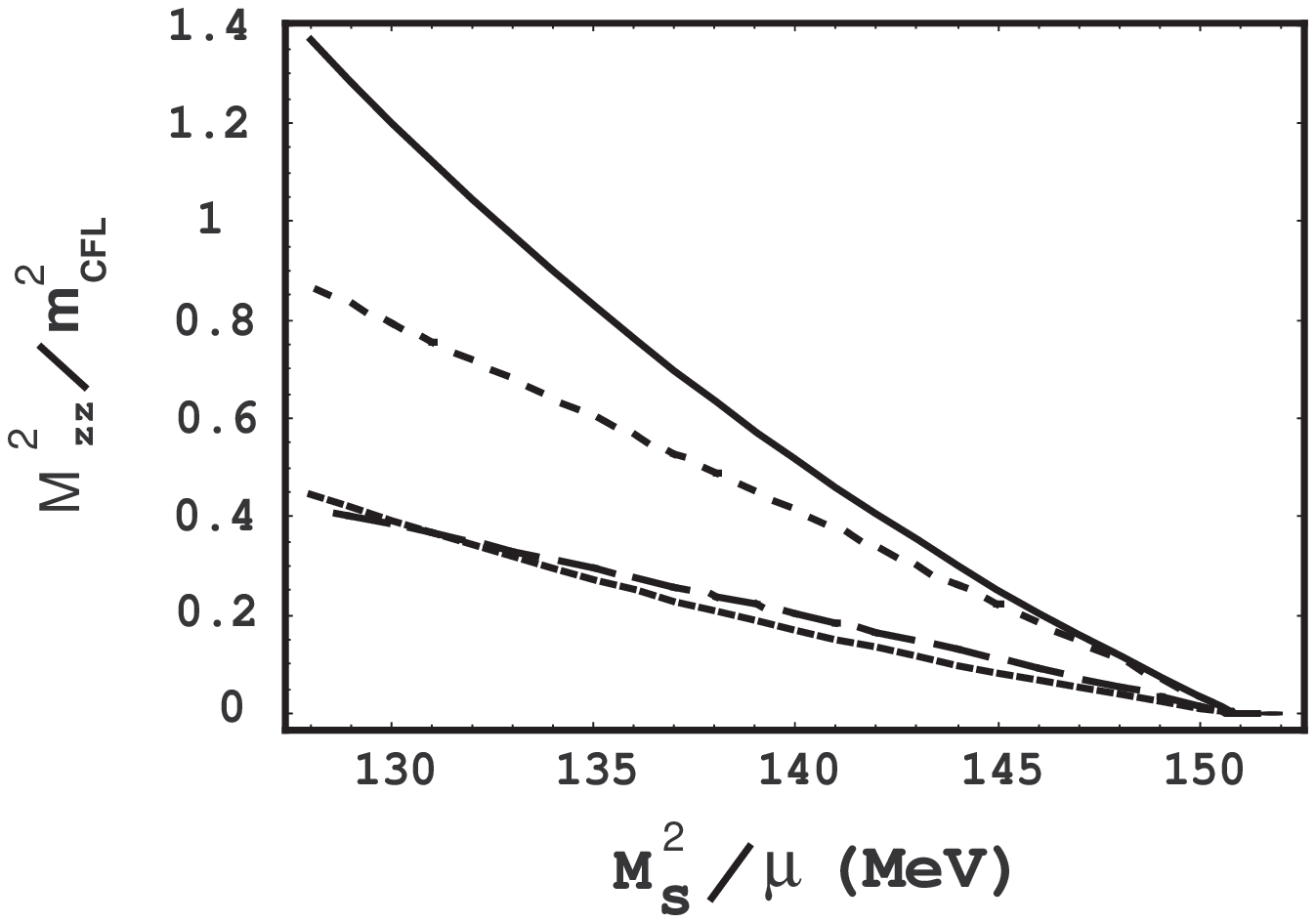}
\includegraphics[width=7cm]{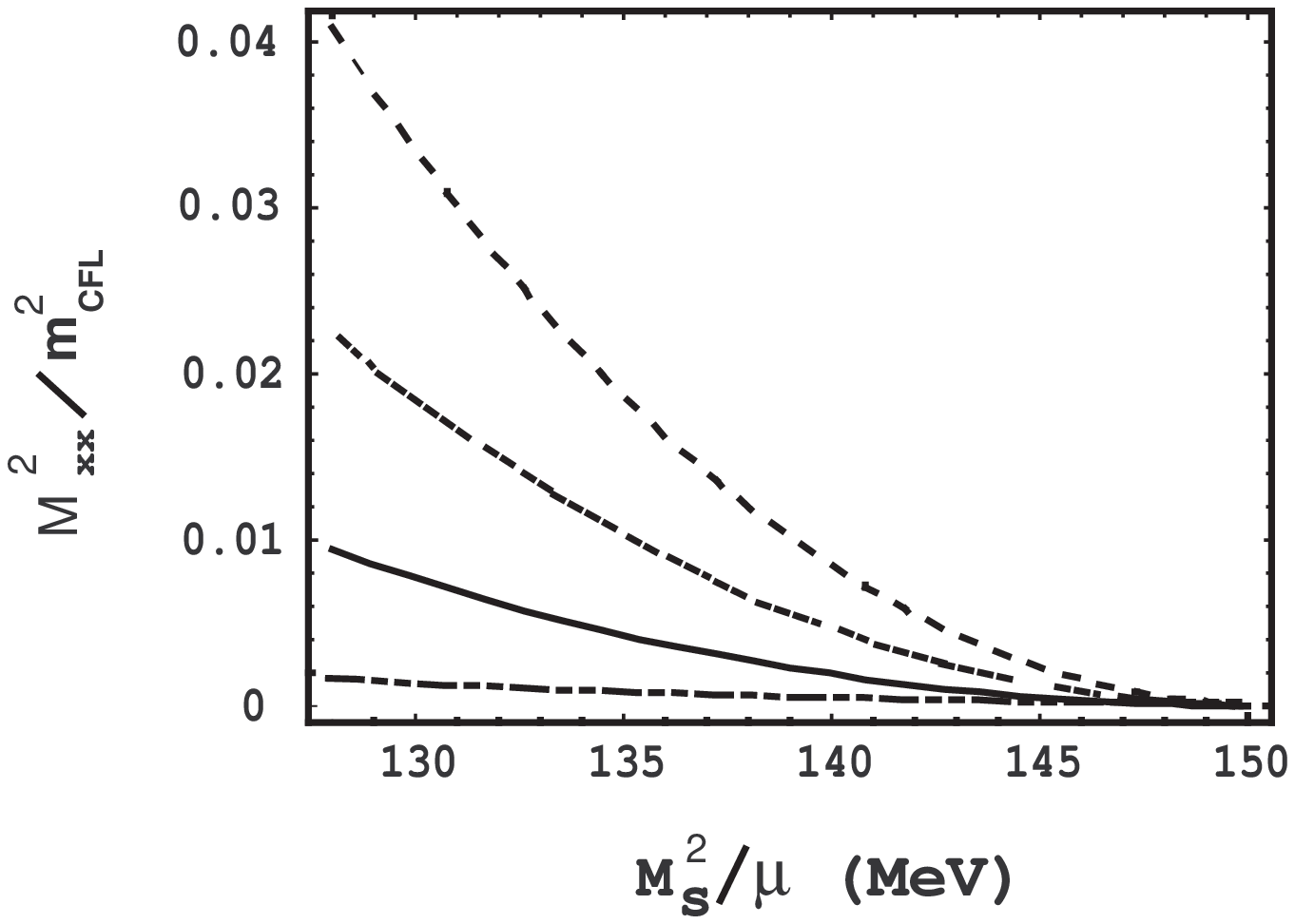}}
\caption{\label{masselong}{ \rm On the top: Longitudinal squared
Meissner masses, in units of the CFL Meissner mass, vs $M_s^2 /\mu$;
from top to bottom the lines  refer to the gluons $A_3$, $A_6=A_7$,
$A_8$ (long dashed), and $A_1=A_2=A_4=A_5$ (dotted line). On the
bottom: Transverse squared Meissner masses; from top to bottom the
lines refer to the gluons $A_6=A_7$, $A_1=A_2=A_4=A_5$, $A_3$, and
$A_8$.}}
\end{figure}

The problem which comes next in the analysis of three flavor LOFF
is then to find out whether such a phase is chromomagnetically
stable. This analysis has been performed for the three flavor LOFF
phase with single plane wave structure \cite{Ciminale:2006sm}, and
the results are shown in Fig.\ref{masselong}. The plotted
quantities are the squared  screening Meissner masses of the
gluons as functions of $m^2_s/\mu$. They are tensors with
longitudinal and transverse components and a nontrivial color
structure, defined as the opposite of the spatial component of the
polarization tensor in the static limit

\begin{equation}
\left({\cal M}^2\right)^{ij}_{ab} \equiv - \Pi_{ab}^{ij}(p_0=0,{\bf
p}=0)~,
\end{equation}
where (i, j) are spatial indices and (a,b) are adjoint color
indices. The gluons Meissner masses are evaluated in
Ginzburg-Landau approximation to the order $\Delta^4$.

All the masses are positive,  hence the three flavor LOFF phase is
chromo-magnetically stable in the Ginzburg-Landau limit. Furthermore
the 3-flavor LOFF phase develops the Meissner screening effect along
the q-direction, being the transverse Meissner masses suppressed as
$\Delta^2/\delta\mu^2$ in comparison with the longitudinal ones.
It's worth to note here that although we have computed the Meissner
masses only for the single plane wave Fulde-Ferrell (FF) structure,
we know from the two flavor case that more complicated crystalline
structures have a lower free energy than the FF state and the same
is true  in the three flavor case, as we will show in the next
section. In the general case one should replace~\eqref{eq:1Ws} with
\begin{equation}
\Delta(r) = \sum_{m=1}^{N}\sum_{I=1}^{3}\Delta_I\exp\left\{{\vec
q}_I~^m\cdot{\vec r}\right\}\epsilon_{ij I}\epsilon^{\alpha\beta
I}
\end{equation}
where ${\vec q}_I~^m$ $(m=1,\dots,N)$ are the momenta which define
the LOFF crystal relative to the condensate $\Delta_I$; the
geometry of the structure and the number $N$ of plane waves should
be determined by minimization of the free energy. Once the optimal
structure is found, one should compute the Meissner masses. If
this structure contains at least three linearly independent
momenta, the Meissner tensor should be positive definite for small
values of $\Delta$, since it is additive with respect to different
terms of~\eqref{eq:1Ws} to order
$\Delta^2$~\cite{Giannakis:2005vw,Gatto:2007ja}. These
considerations suggest that a LOFF crystal can remove the
chromo-magnetic instability of the homogeneous superconductive
phases of QCD, resulting as the true vacuum of the theory.

\subsection{Cooling and neutrino emission of a compact star with LOFF matter core}

\begin{figure}[h!] \centering
{\includegraphics[width=7cm,angle=0.0]{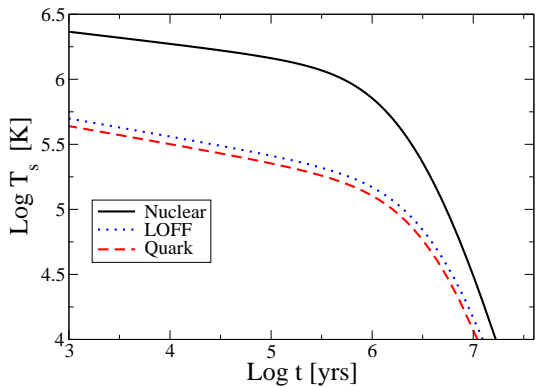}}
\caption{\label{coolsurf}{ \rm Surface temperature
$T_s$, in Kelvin, as a function of time, in years, for the three toy
models of compact stars described in the text.}}
\end{figure}

In \cite{Anglani:2006br} the specific heat and neutrino emissivity
due to direct URCA processes for quark matter in the color
superconductive LOFF state with single plane wave pairing are
evaluated. The resulting surface cooling curves are shown in Fig.
\ref{coolsurf}. The three lines included in this figure
correspond to three different toy stellar models studied in
\cite{Anglani:2006br}. The solid line (black online) describes
 the cooling of a compact star made of electrically neutral nuclear matter of non interacting neutrons, protons and electrons in beta equilibrium; the dashed
curve (red online) refers to a toy star with nuclear matter mantle
and a core of unpaired quark matter, interacting {\it via} gluon
exchange; the dotted line (blue online) is for a nuclear matter
mantle and a core of quark matter in the LOFF state. This last model
is computed for $\mu=500$ MeV and $m^2_s/\mu=140$ MeV. This results, that should be considered preliminary, apparently show that a toy star with a LOFF matter core seems to cool down faster than an ordinary neutron star. This might have interesting phenomenological consequences.

\section{Three flavor crystalline phases}

The choice of keeping the wave vectors on the same direction is the
simplest, but not necessarily the favored one. This is because the
minimization of the thermodynamic potential with respect to the norm
of the wave vectors does not give any information about the
orientation and the eventual crystalline structure chosen by the
system. The only way to determine it is to compare different
possible structures and choose the one with lower free energy.

\begin{figure}[t]
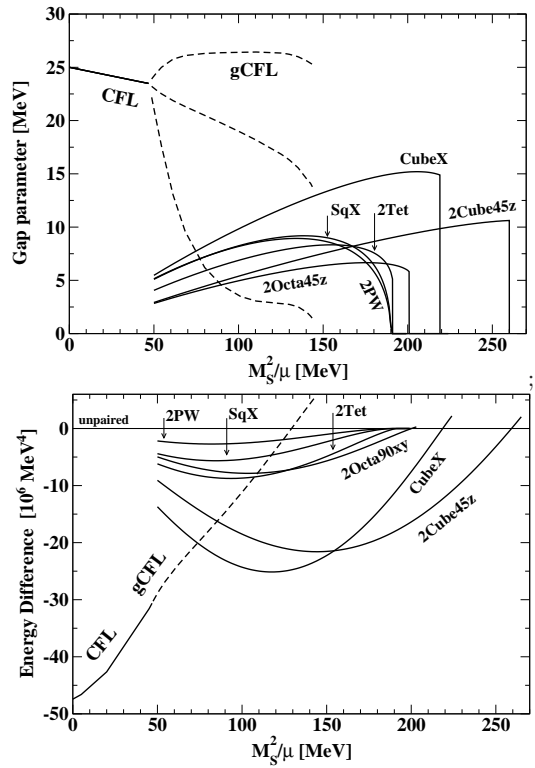
 \centering
{\includegraphics[width=2.7in,angle=0]{dtakr7.eps};
\includegraphics[width=2.7in,angle=0]{omegakr6.eps}}
\caption{Gap parameter and Free energy versus $M_s^2/\mu$ for
three-flavor crystalline color superconducting phases with various
crystal structures.  The crystal structures are described in
\cite{Rajagopal:2006ig}. For comparison, the CFL and gCFL
corresponding parameters are shown, as found in
\cite{gCFL}. The plots are taken from
\cite{Rajagopal:2006ig}.} \label{deltavsx}
\end{figure}

The crystallographic study performed in \cite{Rajagopal:2006ig} with
a Ginzburg-Landau expansion up to the sixth order in the order
parameter $\Delta$ reveals two very robust structures, the CubeX and
2Cube45z. Their composition is explained in \cite{Rajagopal:2006ig}.
Figure \ref{deltavsx} shows that they have gaps and free energies as
large as one half of corresponding CFL parameters. The figure on the top reveals a first order transition to the normal phase for both the favored structures, within the Ginzburg-Landau analysis,
while the figure on the bottom shows that the window in $M^2_s/\mu$, where the
free energy of at least one of these structures is lower than the
normal and the other color superconducting phases, is broad, e.g.

\begin{equation}
72 \; \mbox{MeV} < m^2_s/\mu < 259\; \mbox{MeV}\,.\label{windowmsmu}
\end{equation}
 The presence of a first order phase transition, starting from a GL expansion in the order
parameter, and the large values for this last one yield to believe
that the quantitative results cannot be considered fully reliable.
Nevertheless the qualitative picture should be correct, indicating
the cubic structures are the preferred ones. Furthermore in the
first work of \cite{Mannarelli:2006fy} it is shown that the GL
analysis tends to underestimate the magnitude of the gap, at least
for the one plane wave state.

\subsection{Self-consistent computation of quark masses}

\begin{figure}[h!] \centering
{\includegraphics[width=7cm,angle=0.0]{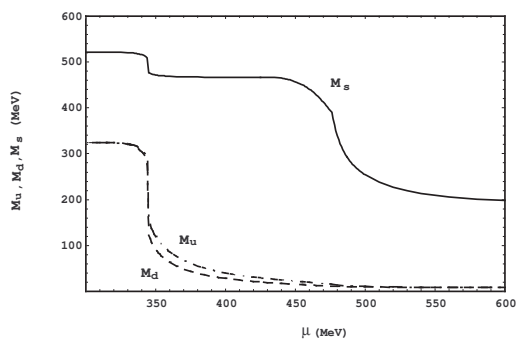}}
\caption{\label{selfcon} Constituent quark masses $M_s$,  $M_d$ and $M_u$  as functions of
the baryon chemical potential $\mu$.}
\end{figure}

The study performed in some works like \cite{Abuki:2004zk} for
normal and homogenous superconducting phases has been extended for
3-flavor LOFF phases in \cite{Ippolito:2007uz}, even if a location
for these phases into the phase diagram of QCD is not yet possible,
since the calculations for these phases are only performed at $T=0$.
The starting point in \cite{Ippolito:2007uz} is a NJL Lagrangean
with 4-fermion and 6-fermion quark-antiquark couplings, plus the
diquark terms for CubeX and 2Cube45z crystalline solutions. Treating
both the quark-quark and the quark-antiquark interaction in a
mean-field approximation, and minimizing the free energy with
respect to the chiral ($<\bar\psi \psi>$) and the diquark ($<\psi
\psi>$) condensates, it is possible to obtain the $\mu$-dependence
of the constituent masses of the quarks, where $\mu$ is the average
chemical potential of the three flavors. The corresponding plot is presented in
Fig.~\ref{selfcon}. Taking in particular the values
for the strange quark constituent mass (the constituent masses of
\emph{up} and \emph{down} quarks are basically the same as their
current masses in the region of interest, since the corresponding
chiral symmetry has already been restored), and putting them in the
window \eqref{windowmsmu}, it is possible to directly obtain the
range of $\mu$ where the crystalline LOFF phases are favored. This
window amounts to be quite large, namely $442 \;\mbox{MeV} <\mu<
515\; \mbox{MeV}\,.$

\subsection{Low energy effective action and masses of pseudo-Goldstones}
The breaking of the symmetry group of QCD
\begin{equation}
\mbox{SU(3)}_c \otimes \mbox{SU(3)}_V \otimes\mbox{SU(3)}_A
\otimes\mbox{U(1)}_V \otimes\mbox{U(1)}_A
\end{equation}
in presence of color superconductivity produces a certain number of
Goldstone bosons, depending on which symmetries are broken by the
particular diquark condensate. In the CFL phase the group is
spontaneously broken to the diagonal $\mbox{SU(3)}_{c+L+R}$ that
locks together the color and flavor symmetries, so giving the name
to the phase. In the case of three flavor crystalline LOFF phases
\cite{Anglani:2007aa} the color gauge group is again spontaneously
broken, so the eight gluons acquire mass by the Anderson-Higgs
mechanism, but the presence of nonzero quark masses in the
Lagrangian and the difference in chemical potential between the
flavors give rise to eight Goldstone modes corresponding to the
$\mbox{SU(3)}_A$ breaking. The breaking of $\mbox{U(1)}_V$ generates
another Goldstone boson, the so-called \emph{superfluid} mode, while
it is assumed that the $\mbox{U(1)}_A$ is restored for the high
values of baryon chemical potential considered therein. In
\cite{Anglani:2007aa} the masses and the decay constants for these
nine mesons are computed, using a Ginzburg-Landau expansion of the
quark propagator up to the second power of $\Delta/\delta\mu$. The
main goal of this paper is to study the possibility of meson
condensation in LOFF phase, as it occurs in CFL phase. The results
at the order $\Delta^2/\delta\mu^2$ show instead that the squared
mass matrix for the octet sector is always definite positive, while
the superfluid mode is massless. This results exclude the
possibility of meson condensation in the LOFF phases with three
flavor, but also indicate that the superfluid mode could be important for
the low energy spectrum of these phases.

%\begin{theacknowledgments}

%\end{theacknowledgments}

%%%%%%%%%%%%%%%%%%%%%%%%%%%%%%%%%%%%%%%%%%%%%%%%
%% The bibliography can be prepared using the BibTeX program or
%% manually.
%%
%% The code below assumes that BibTeX is used.  If the bibliography is
%% produced without BibTeX comment out the following lines and see the
%% aipguide.pdf for further information.
%%
%% For your convenience a manually coded example is appended
%% after the \end{document}
%%%%%%%%%%%%%%%%%%%%%%%%%%%%%%%%%%%%%%%%%%%%%%%%

%%%%%%%%%%%%%%%%%%%%%%%%%%%%%%%%%%%%%%%%%%%%%%%%
%% You may have to change the BibTeX style below, depending on your
%% setup or preferences.
%%
%%
%% For The AIP proceedings layouts use either
%%%%%%%%%%%%%%%%%%%%%%%%%%%%%%%%%%%%%%%%%%%%

%\bibliographystyle{aipproc}   % if natbib is available
%\bibliographystyle{aipprocl} % if natbib is missing

%%%%%%%%%%%%%%%%%%%%%%%%%%%%%%%%%%%%%%%%%%%
%% You probably want to use your own bibtex database here
%%%%%%%%%%%%%%%%%%%%%%%%%%%%%%%%%%%%%%%%%%%
%\bibliography{Proceeding}

%%%%%%%%%%%%%%%%%%%%%%%%%%%%%%%%%%%%%%%%%%%
%% Just a reminder that you may have to run bibtex
%% All of it up to \end{document} can be removed
%% if you don't like the warning.
%%%%%%%%%%%%%%%%%%%%%%%%%%%%%%%%%%%%%%%%%%%
%\IfFileExists{\jobname.bbl}{}
 %{\typeout{}
  %\typeout{******************************************}
  %\typeout{** Please run "bibtex \jobname" to optain}
  %\typeout{** the bibliography and then re-run LaTeX}
  %\typeout{** twice to fix the references!}
  %\typeout{******************************************}
  %\typeout{}
% }

\end{document}